\documentclass[prd,twocolumn,showpacs,floatfix,amsmath,nofootinbib,amssymb,floatfix]{revtex4}
\usepackage{graphicx,color,dcolumn,booktabs,bm}
\usepackage{longtable,lscape}
\usepackage{pdfpages}
\usepackage{txfonts}
\usepackage{overpic}
\usepackage{amssymb}
\usepackage{indentfirst}
\usepackage{feynmf}   
\usepackage{slashed}  
\usepackage{cases}
\usepackage{color}
\usepackage{multirow}
\usepackage{threeparttable}
\usepackage{epstopdf}
\usepackage{enumerate}
\usepackage{graphicx,color,dcolumn,booktabs,bm}
\usepackage[colorlinks, citecolor=blue,anchorcolor=red,menucolor=red, linkcolor=red,filecolor=red,urlcolor=blue,frenchlinks=red]{hyperref}

\graphicspath{{Figures/}} %

\begin{document}

\title{$K(1690)$ signal from COMPASS as a strange hybrid state}

\author{Bing Chen$^{1,3}$}\email{chenbing@ahstu.edu.cn}
\author{Ri-Qing Qian$^{2,3,4,5}$}\email{qianrq@lzu.edu.cn}
\author{Xiang Liu$^{2,3,4,5}$}\email{xiangliu@lzu.edu.cn}
\affiliation{ $^1$School of Electrical and Electronic Engineering, Anhui Science and Technology University, Bengbu 233000, China\\
$^2$School of Physical Science and Technology, Lanzhou University, Lanzhou 730000, China\\
$^3$Lanzhou Center for Theoretical Physics,
Key Laboratory of Theoretical Physics of Gansu Province,\\
Key Laboratory of Quantum Theory and Applications of MoE,\\
Gansu Provincial Research Center for Basic Disciplines of Quantum Physics, Lanzhou University, Lanzhou 730000, China\\
$^4$MoE Frontiers Science Center for Rare Isotopes, Lanzhou University, Lanzhou 730000, China\\
$^5$Research Center for Hadron and CSR Physics, Lanzhou University and Institute of Modern Physics of CAS, Lanzhou 730000, China}

\date{\today}

\begin{abstract}

A pseudoscalar resonance structure, denoted as the $K(1690)$, was recently discovered by the COMPASS Collaboration in the scattering reaction $K^-+p\to K^-\pi^-\pi^++p$. If the $K(1690)$ is a genuine state, there are three observed pseudoscalar strange mesons, namely the $K(1460)$, $K(1690)$, and $K(1830)$, in the 1.0$-$2.0 GeV region. However, within the conventional quark model, only the $2^1S_0$ and $3^1S_0$ strange meson states are expected in this energy region. Therefore, at least one of these states should be interpreted as an exotic candidate. In this work, we study the spectrum of strange mesons systematically, and find that the $K(1460)$ and $K(1830)$ are good candidates of $2^1S_0$ and $3^1S_0$ strange mesons, respectively. A further investigation of their strong decays also supports this assignment. Furthermore, we note that the production mechanism of the $K(1690)$ is similar to that of the $\pi_1(1600)$, which has been widely regarded as a hybrid meson candidate. We investigate the strong decays of the $K(1690)$ within a constituent gluon model by treating it as a $0^-$ strange hybrid meson and find that the results support the $K(1690)$ as a hybrid state. Finally, we identify several important decay modes of the $K(1690)$ that may be used in future experiments to test whether it contains the $n^1S_0$ ($n=$2 and 3) $s\bar{q}$ component. We also suggest BESIII to search for the $K(1690)$ state through the $J/\psi\to K K(1690) \to KK^{\ast}_0(1430)\pi\to KK\pi\pi$ process.

\end{abstract}

%


\maketitle

\section{Introduction}\label{sec1}

Since a complete understanding of all strong-interaction phenomena cannot yet be derived directly from quantum chromodynamics (QCD), the phenomenological models have long played an essential role in hadron physics. Among them, the conventional quark model has provided a successful framework for describing the properties of meson and baryon systems. However, experiments over the past two decades have observed numerous hadronic states that cannot be readily accommodated within the conventional quark model~\cite{Chen:2016qju,Guo:2017jvc,Olsen:2017bmm,Liu:2019zoy,Brambilla:2019esw,Chen:2022asf,Bai:2026atm}. These states are possible candidates for the exotic hadronic states. 


Although many new resonances have been observed experimentally, identifying an exotic state theoretically remains challenging.
In some cases, a resonance can be considered exotic if it possesses unusual quantum numbers that cannot be accommodated within the conventional quark model. For example, the $\pi_1(1600)$ with exotic quantum numbers $J^{PC}=1^{-+}$ has been widely regarded as a hybrid meson~\cite{Benhamida:2019nfx,Eshraim:2020ucw,Shastry:2022mhk,Chen:2023ukh}.
In other cases, when a resonance state has the same quantum numbers as the conventional hadrons, the identification of exotic state becomes more complicated because it must be carefully distinguished from the conventional hadrons. However, if the number of observed states with a particular $J^{P(C)}$ quantum number exceeds the expectation of the conventional quark model, these additional states may provide relatively clear candidates for the exotic hadrons.

Recently, the COMPASS Collaboration observed a resonance structure around 1.7 GeV in the $0^-0^+K\rho (770)$ wave~\cite{COMPASS:2025wkw}. In the same cascade process
\begin{equation}
K^-+p\to\rho^0K^-+p \to \pi^+\pi^-K^-+p, \label{eq1}
\end{equation}
they also found the $K(1460)$ and $K(1830)$ states. According to the results of quark potential model \cite{Godfrey:1985xj,Vijande:2004he,Pang:2017dlw,Taboada-Nieto:2022igy} and Regge phenomenology \cite{Chen:2018nnr,Oudichhya:2023lva}, only the $2^1S_0$ and  $3^1S_0$ strange mesons could appear in the mass region 1.0$-$2.0 GeV. Here, the strange mesons refers specifically to the $s\bar{u}$, $s\bar{d}$, $\bar{s}u$ and $\bar{s}d$ states. Therefore, among the $K(1460)$, $K(1690)$, and $K(1830)$ states, there should exist a supernumerary state that cannot be assigned as a strange meson. By further comparing the measured masses with the theoretical results, the $K(1460)$ and $K(1830)$ could be the $2^1S_0$ and  $3^1S_0$ mesons, respectively. Then the $K(1690)$ becomes a supernumerary state in the conventional quark model because it has no position in the strange meson family. In fact, the $K(1690)$ has been regarded as a crypto-exotic state by the COMPASS Collaboration \cite{COMPASS:2025wkw}. The resonance parameters of $K(1460)$, $K(1690)$, and $K(1830)$ are listed in Table \ref{table1}.

\begin{table}[htbp]
\caption{The measured masses and decay widths of $K(1460)$, $K(1690)$, and $K(1830)$ (in units of MeV).} \label{table1}
\renewcommand\arraystretch{1.2}
\begin{tabular*}{86mm}{@{\extracolsep{\fill}}ccc}
\toprule[1pt]\toprule[1pt]
  State      &  Mass     &    Width  \\
\toprule[1pt]
  $K(1460)$~\cite{LHCb:2017swu}     &  $1482.40\pm3.58\pm15.22$  &  $335.60\pm6.20\pm8.65$  \\
  $K(1690)$~\cite{COMPASS:2025wkw}  &  $1687\pm10^{+2}_{-67}$     &  $140\pm20^{+50}_{-50}$   \\
  $K(1830)$~\cite{COMPASS:2025wkw}  &  $1893\pm17^{+13}_{-39}$   &   $160\pm40^{+60}_{-80}$    \\
\bottomrule[1pt]\bottomrule[1pt]
\end{tabular*}
\end{table}

In this work, we examine the assignments of the $K(1460)$ and $K(1830)$ as the $2^1S_0$ and $3^1S_0$ $K$-meson states, respectively. Meanwhile, we note that diffractive $Kp$ reactions may provide a favorable environment for producing hybrid mesons. For instance, the $1^{-+}$ $\pi_1(1600)$ state, observed in the $\pi^- p \to \pi^+\pi^-\pi^- p$ process \cite{E852:1998mbq,COMPASS:2009xrl,COMPASS:2021ogp}, has been interpreted as a hybrid meson \cite{Benhamida:2019nfx,Eshraim:2020ucw,Shastry:2022mhk,Chen:2023ukh}. Motivated by these considerations, we investigate the possibility that the $K(1690)$ is a hybrid state. We adopt the unified model parameters to describe both the strange meson and the $K$ hybrid, following the approach of Ref.~\cite{Chen:2023ukh}. In this way, we can not only address the puzzle of the three pseudoscalar $K$ states in the 1--2~GeV region, but also provide useful clues for future experiments to further test the nature of the $K(1690)$.

The paper is organized as follows. In Sec. \ref{sec2}, we examine the assignments of the $K(1460)$ and $K(1830)$ to the $2^1S_0$ and $3^1S_0$ states, respectively, where their masses and strong decay properties are investigated.
Assuming the $K(1690)$ to be a pseudoscalar hybrid meson, we study its strong decays within the constituent gluon model in Sec. \ref{sec3}. Then, the possible mixing of the $K(1690)$ with the $n^1S_0$ ($n=$2 and 3) $s\bar{q}$ component is discussed. Finally, the paper ends with a short summary in Sec. \ref{sec5}.

\section{The $K(1460)$ and $K(1830)$ as $2^1S_0$ and $3^1S_0$ states}\label{sec2}

First, we study the spectrum of $K$ mesons using the quark potential model to testing the assignment of $2^1S_0$ and $3^1S_0$ of the $K(1460)$ and $K(1830)$, respectively. In  our calculations, the following spinless Salpeter equation is employed:
\begin{equation}
\left[\sqrt{m_1^2+p^2}+\sqrt{m_2^2+p^2}+V(r)\right]\psi_{nL}=E\psi_{nL}. \label{eq2}
\end{equation}
The potential $V(r)$ is the usual Cornell potential with spin-dependent corrections:
\begin{equation}
  V(r)=V_0(r) + V'(r) \,,
\end{equation}
where
\begin{eqnarray}
  V_0(r) &=& -\frac{4}{3}\frac{\alpha}{r}+br+c+\frac{32\alpha\sigma^3e^{-\sigma^2r^2}}{9\sqrt{\pi}m_1m_2}\textbf{s}_1\cdot\textbf{s}_2 \,,\\
  V'(r) &=&  \frac{4\alpha}{3r^3}\frac{\textbf{S}\cdot\textbf{L}+\hat{\texttt{S}}_{12}}{m_1m_2}+\left(\frac{2\alpha}{3r^3}-\frac{b}{2r}\right)\left(\frac{\textbf{s}_1}{m_1^2}+\frac{\textbf{s}_2}{m_2^2}\right)\cdot\textbf{L}\,.
\end{eqnarray}
This model has been applied to calculate the mass spectrum of light flavor mesons in our previous work \cite{Chen:2023ukh}, where the masses of $1S$ and $2S$ $K$ mesons have been predicted. In the present work, we extend the calculation to obtain the complete mass spectrum of $K$ mesons, including the $3^1S_0$ state. We adopt the same model parameters as those in Ref. \cite{Chen:2023ukh}, i.e., $\alpha=0.64$, $b=0.165$ GeV$^2$, $m_{u,d}=0.32$ GeV, $m_s=0.45$ GeV, $c=-0.398$ GeV, $\sigma=0.45$ GeV.

We solve Eq.~\eqref{eq2} with potential $V_0$ using the Gaussian expansion method, and then the potential $V'$ are treated as perturbations to obtain the mass spectrum of $K$ mesons. The resulting spectrum for the $K$ mesons is shown in Table~\ref{table2}, together with their experimental candidates.

Due to the mass difference of $u/d$ and $s$ quarks, the $\boldsymbol{L}\cdot \boldsymbol{s}_i$ term will induce mixing between $n^1L_L$ and $n^3L_L$ states. The mixed states, which are labeled as $n L_L$ and $nL^\prime_{L}$, can be expressed as
\begin{eqnarray}
\begin{aligned}
 \left(
           \begin{array}{c}
                    nL_{J=L} \\
                    nL^\prime_{J=L} \\
                    \end{array}
     \right)&=\left(
           \begin{array}{cc}
                    \cos \theta   & -\sin\theta   \\
                    \sin\theta    & ~~\cos\theta \\
                    \end{array}
     \right)  \left(
           \begin{array}{c}
                     n^{3}L_{J=L} \\
                     n^{1}L_{J=L} \\
                    \end{array}
     \right). \label{eq6}
\end{aligned}
\end{eqnarray}
The $nL_{J=L}$ state corresponds to the $K$ meson with heavier mass, while the $nL^\prime_{J=L}$ refers to the $K$ meson with lighter mass. The mixing angles are given for the physical $K$ mesons with $J=L$ [$(J\geq1)$] in Table~\ref{table2}.
\begin{table}[htbp]
\caption{Predicted masses of the $K$ states (in units of GeV). The mixing angles for the physical $K$ mesons with $J=L$ ($(J\geq1)$) are $\theta_{1P}=26.6^\circ$, $\theta_{2P}=20.8^\circ$, $\theta_{3P}=19.3^\circ$, $\theta_{1D}=30.3^\circ$, $\theta_{2D}=29.8^\circ$,  $\theta_{1F}=32.3^\circ$,  $\theta_{2F}=40.9^\circ$, and  $\theta_{1G}=33.7^\circ$, respectively.
}\label{table2}
\renewcommand\arraystretch{1.15}
\begin{tabular*}{86mm}{l@{\extracolsep{\fill}}cclcc }
\toprule[1pt]\toprule[1pt]
 State    & Pred$.$      &   Exp$.$ \cite{ParticleDataGroup:2024cfk}   &   State         &  Pred$.$        &   Exp$.$ \cite{ParticleDataGroup:2024cfk}   \\
\toprule[1pt]
  $1^1S_0$        &   0.492      &   $K(496)$       &  $1^3P_0$        &   1.418      & $K^\ast_0(1430)$ \\
  $1^3S_1$        &   0.899      &   $K^\ast(892)$  & $1P^\prime_1$    &   1.271      &   $K_1(1270)$    \\
  $2^1S_0$        &   1.344      &  $K(1460)$       &  $1P_1$          &   1.483      &   $K_1(1400)$    \\
  $2^3S_1$        &   1.531      &  $K^\ast(1410)$  &  $1^3P_2$        &   1.384      & $K^\ast_2(1430)$ \\
  $3^1S_0$        &   1.871      &   $K(1830)$      &  $2^3P_0$        &   1.952      & $K^\ast_0(1950)$ \\
  $3^3S_1$        &   2.011      &   $\cdots$       & $2P^\prime_1$    &   1.786      &   $\cdots$       \\
  $4^1S_0$        &   2.297      &   $\cdots$       &  $2P_1$          &   1.977      &   $\cdots$       \\
  $4^3S_1$        &   2.417      &   $\cdots$       &  $2^3P_2$        &   1.870      & $K^\ast_2(1980)$ \\
  $3^3P_0$        &   2.364      &   $\cdots$       &  $2^3D_1$        &   2.301      &    $\cdots$      \\
 $3P^\prime_1$    &   2.215      &   $\cdots$       & $2D^\prime_2$    &   2.111      &    $\cdots$      \\
  $3P_1$          &   2.379      &   $\cdots$       &  $2D_2$          &   2.251      &    $K_2(2250)$   \\
  $3^3P_2$        &   2.279      &   $\cdots$       &  $2^3D_3$        &   2.113      &    $\cdots$      \\
  $1^3D_1$        &   1.907      &   $\cdots$       &  $1^3F_2$        &   2.225      &    $\cdots$      \\
 $1D^\prime_2$    &   1.695      &   $K_2(1770)$    & $1F^\prime_3$    &   2.012      &    $\cdots$      \\
  $1D_2$          &   1.847      &   $K_2(1820)$    &  $1F_3$          &   2.141      &    $\cdots$      \\
  $1^3D_3$        &   1.662      & $K^\ast_3(1780)$ &  $1^3F_4$        &   1.929      &    $K^\ast_4(2045)$ \\
  $2^3F_2$        &   2.567      &   $\cdots$       &  $1^3G_3$        &   2.493      &    $\cdots$      \\
 $2F^\prime_3$    &   2.385      &   $K_3(2320)$    & $1G^\prime_4$    &   2.278      &    $\cdots$      \\
  $2F_3$          &   2.502      &   $\cdots$       &  $1G_4$          &   2.399      &    $K_4(2500)$   \\
  $2^3F_4$        &   2.335      &   $\cdots$       &  $1^3G_5$        &   2.177      &    $K^\ast_5(2380)$ \\
\bottomrule[1pt]\bottomrule[1pt]
\end{tabular*}
\end{table}

As shown in Tables \ref{table2}, the measured masses of most discovered $K$ mesons could be fairly reproduced by the simple quark potential model. However, the masses of some $n^3L_{J=L\pm1}$ $K$ mesons seem to be underestimated by 100$\sim$200 MeV. The probable reason is that the spin-orbit and tensor interactions in $V'(r)$ are obtained in the nonrelativistic limit. When the relativistic effects are incorporated into the potential model, the predicted precision for the masses of $n^3L_{J=L\pm1}$ $K$ mesons could be improved \cite{Godfrey:1985xj,Pang:2017dlw}. Here, we focus on the pseudoscalar $K$ mesons in the 1.0$-$2.0 GeV region, which are unaffected by the orbit and tensor interactions. According to our result in Table \ref{table2}, the predicted masses favor assigning the $K(1460)$ and $K(1830)$ as the $2^1S_0$ and $3^1S_0$ states, respectively. This assignment is also supported by the results from Refs. \cite{Godfrey:1985xj,Vijande:2004he,Pang:2017dlw,Chen:2018nnr,Taboada-Nieto:2022igy,Oudichhya:2023lva} where the mass of $3^1S_0$ strange meson was predicted to be about 1.90$\sim$2.0 GeV. Thus, the observed $K(1690)$ could not be a conventional strange meson since its mass is light for the $3^1S_0$ state, but heavy for the $2^1S_0$ state.


To further test the assignment of $K(1460)$ and $K(1830)$ as the $2^1S_0$ and $3^1S_0$ strange mesons, respectively, we study their strong decays within the $^3P_0$ model \cite{Micu:1968mk,LeYaouanc:1972vsx}. In this model, the transition matrix element for the process $A\to B+C$ is written as $\langle{BC}|\mathcal {\hat{T}}_{\textup{QPC}}|A\rangle=\delta^3(\textbf{\textrm{K}}_B+\textbf{\textrm{K}}_C)\mathcal {M}^{j_A,j_B,j_C}(p)$, where the transition operator $\mathcal {\hat{T}}_{\textup{QPC}}$ reads as
\begin{equation}
\begin{split}
\mathcal {\hat{T}}_{\textup{QPC}}=&-3\gamma
\sum_{\text{\emph{m}}}\langle1,m;1,-m|0,0\rangle \iint
d^3\textbf{\textrm{k}}_{\mu}d^3\textbf{\textrm{k}}_{\nu}\delta^3(\textbf{\textrm{k}}_{\mu}+\textbf{\textrm{k}}_{\nu})\\ &\times\mathcal
{Y}_1^m\left(\frac{\textbf{\textrm{k}}_{\mu}-\textbf{\textrm{k}}_{\nu}}{2}\right)\omega_0^{({\mu},{\nu})}\varphi^{({\mu},{\nu})}_0\chi^{({\mu},{\nu})}_{1,-m}d^\dag_{{\mu}}(\textbf{\textrm{k}}_{\mu})d^\dag_{{\nu}}(\textbf{\textrm{k}}_{\nu})
\end{split}\label{eq3}
\end{equation}
in the nonrelativistic limit. The $\omega_0^{({\mu},{\nu})}$ and $\varphi^{({\mu},{\nu})}_0$ are the color and flavor wave functions of the $q_{\mu}\bar{q}_{\nu}$ pair created from the vacuum. Explicitly, one has $\omega_0^{({\mu},{\nu})}=(R\bar{R}+G\bar{G}+B\bar{B})/\sqrt{3}$ and $\varphi^{({\mu},{\nu})}_0=(u\bar{u}+d\bar{d}+s\bar{s})/\sqrt{3}$, which denote the color and flavor singlets. The $\chi^{({\mu},{\nu})}_{1,-m}$ represents the pair production in a spin triplet state. When the mock state~\cite{Hayne:1981zy} is adopted to describe the wave function of a hadron state, the partial wave amplitudes $\mathcal{M}_{LS}(p)$ can be obtained by the following formula,
\begin{equation}
\begin{aligned}\label{eq4}
\mathcal {M}^{A\rightarrow B+C}_{LS}(p)=&\frac{\sqrt{4\pi(2L+1)}}{2J_A+1}\sum_{\text{$j_B$,$j_C$}}\langle L0Sj_A|J_Aj_A\rangle\\
&\times\langle J_Bj_B,J_Cj_C|Sj_A\rangle\mathcal
{M}^{j_A,j_B,j_C}(p).
\end{aligned}
\end{equation}
Here, $p$ represents the momentum of an outgoing meson in the rest frame of a meson $A$. The $J_i$ and $j_i$ ($i=$ $A$, $B$, and $C$) denote the total angular momentum and its projection along the $z$ axis of the corresponding hadron states, respectively. $L$ denotes the orbital angular momenta between the final state $B$ and $C$. Finally, the partial width of $A\rightarrow B + C$ can be obtained in terms of the partial wave amplitudes as
\begin{equation}
\begin{aligned}\label{eq5}
\Gamma(A\rightarrow BC)=2\pi\frac{E_BE_C}{M_A}p\sum_{L,S}|\mathcal
{M}^{A\rightarrow B+C}_{LS}(p)|^2
\end{aligned}
\end{equation}
in the $A$ rest frame. In our calculation, the spatial wave functions of the initial and final states are approximated by the simple harmonic oscillator (SHO) wave function. The scale parameter $\beta$ of the SHO wave function is determined by solving the Salpeter equation. The scale parameter $\beta$ of $1S$, $1P$, and $2S$ light mesons have been given in Ref. \cite{Chen:2023ukh}. The $\beta=0.369$ is supplemented for the $3^1S_0$ $K$ meson. More details on the construction of the meson wave functions can be found in our previous works ~\cite{Chen:2016iyi,Chen:2017gnu}.

At present, the total and partial decay widths of $K_2^\ast(1430)$ state have been well measured experimentally \cite{ParticleDataGroup:2024cfk}. In our calculations, the dimensionless parameter $\gamma$ of $^3P_0$ model \footnote{The parameter $\gamma$  describes the strength of the quark-antiquark pair creation from the vacuum.} is fixed by fitting the $K_2^\ast(1430)$ decay widths. In this way, the $K_2^\ast(1430)$ not only serves to determine the parameter $\gamma$ , but also provides a test of $^3P_0$ model.

\begin{table}[htbp]
\caption{Measured and calculated partial decay widths of the $K^\ast_2(1430)$ state (in units of MeV).
}\label{table4}
\renewcommand\arraystretch{1.15}
\begin{tabular*}{86mm}{c@{\extracolsep{\fill}}cccccc}
\toprule[1pt]\toprule[1pt]
 $K^\ast_2(1430)$   & $K\pi$               &   $K^\ast\pi$    &  $K\rho$          &   $K\omega$      &  $K^\ast\pi\pi$   &  Total  \\
\toprule[1pt]
  Expt$.$ \cite{ParticleDataGroup:2024cfk}  &   49.9$\pm$1.2   &   24.7$\pm$1.5    &     8.7$\pm$0.8  &     2.9$\pm$0.8   &     13.4$\pm$2.2  &     $\approx$100.0        \\
  Pred$.$           &   49.8               &   24.0           &     11.7          &     3.2          &     $\cdots$      &      $>$ 88.7      \\
\bottomrule[1pt]\bottomrule[1pt]
\end{tabular*}
\end{table}

The partial and total decay widths of the $K^\ast_2(1430)$ calculated by the $^3P_0$ model are given in Table \ref{table4}, where the experimental results are also listed for comparison. The partial decay width of $K^\ast\pi\pi$ mode has not been calculated. This decay mode may arise from either the direct three-body decay or the cascade process such as $K^\ast_2(1430)\to K^\ast\sigma\to K^\ast\pi\pi$. Since the effectiveness of the $^3P_0$ model for depicting the three-body decay process has never been seriously examined, we restrict our analysis to the two-body OZI-allowed decays, as shown in Table \ref{table4}. It is evident that the decay properties of $K^\ast_2(1430)$ can be well described by the $^3P_0$ model. Finally, the parameter $\gamma=14.7$ is fixed in our calculations. With the parameter $\gamma$, the decays of $K(1460)$ and $K(1830)$ are predicted by the $^3P_0$ model. The concrete results are listed in Table \ref{table5}, where the $K(1460)$ and $K(1830)$ are assigned as the $2^1S_0$ and $3^1S_0$ states, respectively.

\begin{table}[htbp]
\caption{Predicted strong decays of the $K(1460)$ and $K(1830)$ (in units of MeV). Here, the $K(1460)$ and $K(1830)$ are regarded as the $2^1S_0$ and $3^1S_0$ strange mesons, respectively. The experimental total widths of $K(1460)$ and $K(1830)$ \cite{ParticleDataGroup:2024cfk} are listed in the bottom row for comparison.
}\label{table5}
\renewcommand\arraystretch{1.15}
\begin{tabular*}{86mm}{c@{\extracolsep{\fill}}cccccc}
\toprule[1pt]\toprule[1pt]
\multicolumn{2}{c}{$K(1460)~[2^1S_0]$}  & \multicolumn{5}{c}{$K(1830)~[3^1S_0]$} \\
\cline{1-2}\cline{3-7}
 $K^\ast\pi$   & 112.7     & 6.9       &  $K^\ast\eta^\prime$   & 0.0      &  $K^\ast_0(1430)\pi$  &  8.0  \\
 $K^\ast\eta$  & 2.3       & 0.4       &  $K\phi$               & 0.4      &  $f_2(1170)K$         &  2.9  \\
 $K\rho$       & 151.5     & 1.0       &  $K^\ast\rho$          & 67.5     &  $a_2(1320)\pi$       &  2.6  \\
 $K\omega$     & 47.2      & 0.3       &  $K^\ast\omega$        & 22.4     &  $K^\ast_2(1430)\pi$  &  15.3  \\
 $K\pi\pi$     & $\cdots$  & $\cdots$  &  $f_0(1370)K$          & 11.6     &  $K^\ast(1410)\pi$    &  114.2  \\
\cline{1-2}\cline{3-7}
  Total        & $>$313.7  &           &                        &          &                       &  $>$253.5  \\
\multicolumn{2}{l}{$335.60\pm6.20\pm8.65$}   &   &   &  &              \multicolumn{2}{r}{$160\pm40^{+60}_{-80}$}   \\
\bottomrule[1pt]\bottomrule[1pt]
\end{tabular*}
\end{table}

The $K(1460)$ has been seen in the $K^\ast\pi$ and $K\rho$ decay channels \cite{ParticleDataGroup:2024cfk}. These observations are consistent with our result since the $K^\ast\pi$ and $K\rho$ are the main decay modes for the $K(1460)$ (see Table \ref{table5}). The total decay width of the $K(1460)$ is expected to be larger than 313.7 MeV since the three-body decay mode, i.e., the $K(1460)\to K\pi\pi$, has been ignored in our calculation.

As a pure $3^1S_0$ strange meson, the total decay width of the $K(1830)$ is predicted to be larger than 253.5 MeV. Then the relatively broad width of the $K(1830)$ could be understood by our result. Furthermore, it seems more feasable for experiments to search for the $K(1830)$ in the $K\pi\pi\pi$ final-state since the $K^\ast\rho$ and $K^\ast(1410)\pi$ are the main decay channels for the $K(1830)$ state, both of which subsequently cascade to $K\pi\pi\pi$ final state. The partial width of the $K(1830)\to K\rho$ channel is predicted to be 1 MeV, which is not a favorable channel for observing the $K(1830)$, although the COMPASS experiment has observed it in this mode. Further experimental studies of the $K(1830)$ in the $K\pi\pi\pi$ final state are therefore suggested for future measurements.

In summary, the main properties of the $K(1460)$ and $K(1830)$, including their masses and strong decays, could be well explained by the conventional quark model when they are assigned as the $2^1S_0$ and $3^1S_0$ $K$ mesons, respectively. Then the $K(1690)$ becomes a supernumerary state since it cannot find a position in the strange meson family.

\section{The $K(1690)$ as a hybrid state}\label{sec3}

As stated before, the production process of the $K(1690)$ is analogous to that of the $\pi(1600)^-$ state. In our previous work, the $\pi(1600)^-$ has been explained as a hybrid meson \cite{Chen:2023ukh}. So it is no surprising that strange hybrid be produced via $Kp$ diffractive scattering process. Furthermore, the average mass of ground $K^-$ hybrids\footnote{The lowest hybrid multiplet include the $0^{-(+)}$, $1^{-(+)}$, $1^{-(-)}$, and $2^{-(+)}$ $q_1\bar{q}_2g$ states.} has been predicted to be 1852 MeV, and the $0^{-}$ $K^-$ hybrid meson is expected to be the lightest among four ground states \cite{Kalashnikova:1993xb,Dudek:2011bn}. So the predicted mass of strange hybrid is fairly close to that of the $K(1690)$.


In this section, we further test the $0^{-}$ hybrid interpretation of the $K(1690)$ by studying its strong decays.
The constituent gluon (CG) model \cite{Farina:2020slb} is employed to investigate the decay properties of $K(1690)$ state. Recently, we have utilized this model to study the $1^{-+}$, $0^{+-}$, and $2^{+-}$ light hybrid mesons \cite{Chen:2023ukh,Chen:2025pvk}.

In the nonrelativistic limit, the transition operator $\hat{\mathcal{T}}_{\textup{CG}}$ of the constituent gluon model can be expressed as
\begin{equation}
\begin{split}
\hat{\mathcal{T}}_{\textup{CG}} = &g_s \sum_{s,s^\prime,\lambda} \iiint \frac{\textup{d}^3\textbf{\textit{p}}_3\textup{d}^3\textbf{\textit{p}}_4\textup{d}^3\textbf{\textit{k}}_g}{\sqrt{2\omega_g}(2\pi)^6} \delta^3(\textbf{\textit{p}}_3 + \textbf{\textit{p}}_4 - \textbf{\textit{k}}) \\
& \times \frac{\lambda^{c_g}}{2} \phi_0^{(34)} \chi_s^\dagger \bm{\sigma} \tilde{\chi}_{s^\prime} \bm{\epsilon}(\hat{\textbf{k}},\nu) b_3^\dagger(\vec{p}_3) d_4^\dagger(\vec{p}_4) a_{\textbf{k}\nu}^{c_g}(\textbf{\textit{k}}),
\label{eq9}
\end{split}
\end{equation}
which describes the dissociation of a gluon into a quark-antiquark pair. Here, $\lambda^{c_g}$ (${c_g} = 1, 2, \dots, 8$) denote the Gell-Mann matrices. The quantities $\chi_{s^{(\prime)}}$ represent the quark spin wave functions, and $\bm{\epsilon}^\mu(\hat{\textbf{k}},\nu)$ denotes the gluon polarization vector. The flavor wave function $\phi_0^{(34)}$ is $(u\bar{u} + d\bar{d} + s\bar{s})/\sqrt{3}$. The parameter $\omega_g$, representing the effective mass of the constituent gluon \cite{Swanson:1997wy}, is taken to be 0.80 GeV. After constructing the mock states of the initial hybrid meson and final mesons, the transition matrix have the form $\langle{BC}|\hat{\mathcal{T}}_{\textup{CG}}|H\rangle=\delta^3(\textbf{K}_B+\textbf{K}_C)\mathcal{M}^{j_H,j_B,j_C}(p)$. Finally, the partial wave amplitude is given as
\begin{equation}
\begin{aligned}\label{eq10}
\mathcal {M}^{H\to BC}_{LS}(p)=&\frac{\sqrt{4\pi(2L+1)}}{2J_H+1}\sum_{\text{$j_B$,$j_C$}}\langle L0Jj_H|J_Hj_A\rangle\\
&\times\langle J_Bj_B,J_Cj_C|Jj_H\rangle\mathcal
{M}^{j_H,j_B,j_C}(p),
\end{aligned}
\end{equation}
and the partial width of the process $H\to BC$ is given as
\begin{equation}
\begin{aligned}\label{eq11}
\Gamma(H\rightarrow BC)=2\pi\frac{E_BE_C}{M_H}p\sum_{L,S}|\mathcal
{M}_{LS}(p)|^2
\end{aligned}
\end{equation}
in the rest frame of $H$. More details of the constituent gluon model could be found in Refs. \cite{Farina:2020slb,Chen:2025pvk}. When the parameters are taken from Ref. \cite{Chen:2023ukh}, the partial and total decay widths of $K(1690)$ are predicted in Table \ref{table6}.

\begin{table}[htbp]
\caption{The partial and total decay widths of the $K(1690)$ state (in units of MeV). Here, the $K(1690)$ is treated as a pure $0^-$ hybrid meson.} \label{table6}
\renewcommand\arraystretch{1.2}
\begin{tabular*}{86mm}{@{\extracolsep{\fill}}ccccc}
\toprule[1pt]\toprule[1pt]
 $K^\ast\pi$          & $K^\ast\eta$         & $\rho{K}$           & $\omega{K}$         &   $\phi{K}$    \\
 14.5                 & 1.4                  &  0.7                & 0.2                 &    1.5             \\
\cline{4-5}
 $K_0^\ast(1430)\pi$  & $K_2^\ast(1430)\pi$  & $K^\ast(1410)\pi$   & Total               &    Exp.~\cite{COMPASS:2025wkw}   \\
  107.0                & 0.0                 & 9.3                 & 134.6               &   140$\pm$20$^{+50}_{-50}$  \\
\bottomrule[1pt]\bottomrule[1pt]
\end{tabular*}
\end{table}

The predicted total decay width of the $K(1690)$ as a hybrid state is about 134.6 MeV, which is consistent with the experimental measurement~\cite{COMPASS:2025wkw}. This agreement supports the hybrid interpretation of the $K(1690)$.
It is noteworthy that the $K\rho$ channel is predicted to have only a small partial width, although this is the channel in which the $K(1690)$ has been observed by the COMPASS experiment~\cite{COMPASS:2025wkw}. In our calculation, the dominant decay mode of the $K(1690)$ is $K_0^\ast(1430)\pi$, and the $K_0^\ast(1430)$ predominantly decays into $K\pi$ in an $S$-wave. This suggests that the $K(1690)$ could be further investigated via the process
\begin{equation}
K+p \to K(1690)+p \to K_0^\ast(1430)\pi+p \to [K\pi]_S \pi+p.
\end{equation}
Although this reaction process was not included in the partial-wave analysis of the COMPASS experiment~\cite{COMPASS:2025wkw}, a broad resonancelike signal with $J^P=0^-$ in the $[K\pi]_S\pi$ channel was clearly seen around 1.7 GeV~\cite{Wallner:2021kbh}. Further experimental studies of the $K(1690)$ through the above reaction chain are therefore needed.

Finally, we note that a $K(1630)$ state with an undetermined $J^P$ quantum numbers is collected by the Particle Data Group (PDG) \cite{ParticleDataGroup:2024cfk}. This state has only been reported in a single bubble-chamber experiment at CERN \cite{Karnaukhov:1998qq,Karnaukhov:2000zb} through the reaction ``$\pi{p}\to(K_S^0\pi^+\pi^-)+\textup{others}$''. Obviously, the $K(1630)$ needs further confirmation from other experiments.


We have shown that the $K(1460)$ and $K(1830)$ states could be assigned as the pure $2^1S_0$ and $3^1S_0$ strange mesons, while the $K(1690)$ could be interpreted as a pseudoscalar hybrid meson. It is worth noting that mixing may exist between $0^-$ $K$ hybrid state and usual strange meson due to their same $J^P$ quantum numbers. At present, the mixing between the hybrid meson and the conventional meson is still unclear. The $K(1690)$, $K(1460)$ and $K(1830)$ provides a useful opportunity to study possible mixing between hybrid and conventional mesons, which cannot occur in hybrid with exotic quantum number such as $1^{-+}$. Here, we do not attempt to calculate the mixing angle by a concrete model, but instead provide some useful clues for experiments to test whether the $K(1690)$ state contains $2^1S_0$ or $3^1S_0$ $s\bar{q}$ component.
\begin{itemize}
    \item As a pure $s\bar{q}g$ state, the partial widths of decay channels $\rho{K}$ and $\omega{K}$ are relatively small for the $K(1690)$ state. However, the conclusion could be changed when the $K(1690)$ contains the $2^1S_0$ $s\bar{q}$  component. As shown in Table \ref{table5}, the $\rho{K}$ and $\omega{K}$ are the predominant decay channels for the $2^1S_0$ $s\bar{q}$ state. Therefore, the precise measurement of the branching ratio of the $\rho{K}$ and $\omega{K}$ may test whether the $K(1690)$ contains the $2^1S_0$ $s\bar{q}$ admixture.

    \item The decay channels $K^\ast\rho$ and $K^\ast\omega$ are forbidden for the $K(1690)$ if it is a pure pseudoscalar hybrid meson (see Appendix.~\ref{app:selection}). However, these two decay channels are predominant for the $3^1S_0$ strange meson (see Table \ref{table5}). We further check the partial width of $K(1690)\to\rho{K^\ast}$ process, where the $K(1690)$ is treated as a pure $2^1S_0$ or $3^1S_0$ $s\bar{q}$ state. The partial width is obtained to be 6$\sim$13 MeV. Therefore, future experiments may search for the $K(1690)$ in the $\rho{K^\ast}$ to examine whether it contains the $n^1S_0$ ($n=$2 and 3) $s\bar{q}$ component.
\end{itemize}

\begin{figure}[htbp]
\centering
\includegraphics[width=8.4cm,keepaspectratio]{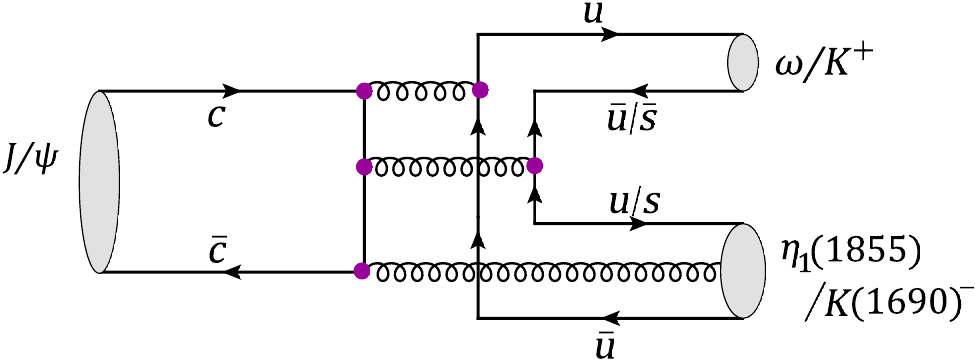}
\caption{A schematic diagram for $J/\psi$ decaying into a light meson ($\omega/K$) and a hybrid state [$\eta_1(1855)/K(1690)$].}
\label{BESIII1690}
\end{figure}

Finally, it is important to search for the $K(1690)$ in other process in addition to the $Kp$ diffractive reaction, as discussed previously.
The annihilation of the $c\bar{c}$ state provides a gluon-rich environment for searching for hybrid meson.
Recently, the BESIII Collaboration observed the hybrid candidate $\eta_1(1855)$ in the $J/\psi\to\omega\eta_1(1855)\to\omega\eta\eta^\prime$ \cite{Liu1855}. The dominant production mechanism of hybrid mesons via the annihilation decay of $J/\psi$ is illustrated in Fig.~\ref{BESIII1690}, and the similar process can also produce a strange hybrid. So, it is promising to observe the $K(1690)$ state through the process of $J/\psi\to KK(1690)\to K K\pi\pi$, which is most feasible on the BESIII experiment.

\section{Summary}\label{sec5}

A pseudoscalar signal, i.e., the $K(1690)$, was recently discovered by the COMPASS Collaboration in the reaction $K^-+p\to K^-\pi^-\pi^++p$ \cite{COMPASS:2025wkw}. The observation of the $K(1690)$ may lead to a puzzle of pseudoscalar $K$ states in the mass region 1.0$-$2.0 GeV. Specifically, three pseudoscalar $K$ states, namely the $K(1460)$, $K(1690)$, and $K(1830)$, have been observed experimentally. However, the conventional quark model allows only the $2^1S_0$ and $3^1S_0$ strange mesons in this region. This indicate that there exist one possible exotic candidate.

In this work, we systematically investigate the spectrum of strange mesons. Our result is consistent with other works \cite{Godfrey:1985xj,Vijande:2004he,Pang:2017dlw,Chen:2018nnr,Taboada-Nieto:2022igy,Oudichhya:2023lva}, supporting the assignment of the $K(1460)$ and $K(1830)$ as the $2^1S_0$ and $3^1S_0$ $K$ mesons. We further test these assignments by investigating their strong decays. We find that the main decay properties of $K(1460)$ and $K(1830)$ could be well described when they are regarded as the $2^1S_0$ and $3^1S_0$ states, respectively. Thus, the $K(1690)$ becomes a supernumerary state within the conventional quark model.

We notice that the production of $K(1690)$ is very similar to that of the $\pi_1(1600)$ \cite{E852:1998mbq,COMPASS:2009xrl,COMPASS:2021ogp}, which has been widely considered being a hybrid candidate. In addition, the measured mass of $K(1690)$ is also in agreement with the expectation of pseudoscalar $s\bar{q}g$ state in our previous work \cite{Chen:2023ukh}. In this work, we further examine the two-body strong decays of $K(1690)$ by a constituent gluon model. We found that the main decay properties of $K(1690)$ could be understood if it is regarded as a pure $0^-$ hybrid meson.

Finally, we compare the main decay channels of $K(1460)$, $K(1690)$, and $K(1830)$ states and provide guidance for probing possible mixing between hybrid and conventional $K$ mesons. We noticed that the $K\rho$ and $K^\ast\rho$ are the main decay modes of $K(1460)$ and $K(1830)$ states, respectively. In contrast, the $K\rho$ is strongly suppressed for $K(1690)$, while the mode $K^\ast\rho$ is a forbidden decay mode. Therefore, we suggest the experiments to search for the $K(1690)$ state in the $K\rho$ and $K^\ast\rho$ channels, which can help to check whether the $K(1690)$ contains the $n^1S_0$ ($n=$2 and 3) $s\bar{q}$ component. In addition, we suggest the BESIII Collaboration to search for the $K(1690)$ in $J/\psi\to{K}{K(1690)}\to{K} K\pi\pi$ process.

\section*{ACKNOWLEDGMENTS}

This work is supported by the Natural Science Foundation of Gansu Province (No. 26RCKA012 and No. 25JRRA799), the National Natural Science Foundation of China under Grants No. 12335001, No. 12247101 and No. 11305003, the ‘111 Center’ under Grant No. B20063, the fundamental Research Funds for the Central Universities (lzujbky-2023-stlt01), and Lanzhou City High-Level Talent Funding. B. C. is also supported by the Talent Attraction Program (DQYJ202003) and the key Discipline Construction Project of Anhui Science and Technology University (No. XK-XJGY002).

\appendix
\section{Selecting rule for $K^*\rho$ and $K^*\omega$ channels}\label{app:selection}
The selection rule, which state that the decay of a pseudoscalar strange hybrid into two $^3S_1$ vector mesons is forbidden,  can be understood as follows.
When the mock states are constructed for the initial and final states in a decay process, the amplitude of $A\to BC$ contains a spin wave function overlap factor
\begin{eqnarray}
\xi_{\mathrm{spin}} = \langle \chi_{S_B M_{S_B}}^{14}\chi_{S_C M_{S_C}}^{32}|\chi_{S_A M_{S_A}}^{12}\chi_{1 m}^{34}\rangle,
\end{eqnarray}
where $\chi_{S_A}^{12}$, $\chi_{S_B}^{14}$, and $\chi_{S_C}^{32}$ denote the spin wave functions of the initial state $A$ and final states $B$ and $C$, respectively. The $\chi_{1}^{34}$ represents the spin wave function of the quark-antiquark pair created from the vacuum (in the $^3P_0$ model) or generated through gluon dissociation (in the constituent gluon model). The quantities $S_A$, $S_B$, and $S_C$ denote the total spins of the initial and final mesons, respectively.
The spin wave function overlap factor can be expressed in terms of the $9-j$ Wigner coefficient as
\begin{eqnarray}\label{eqapp}
  \xi_{\mathrm{spin}} &=& \sum_{S M_S} \langle S_B M_{S_B}, S_C M_{S_C}|S M_S\rangle \langle S_A M_{S_A}, 1 m|S M_S\rangle \nonumber\\
  &&\times \langle (s_1s_4)S_B, (s_3s_2)S_C; S M_S| (s_1s_2)S_A, (s_3s_4)1; S M_S\rangle \nonumber\\
  &=& \sum_{S M_S} \langle S_B M_{S_B}, S_C M_{S_C}|S M_S\rangle \langle S_A M_{S_A}, 1 m|S M_S\rangle \nonumber\\
  &&\times (-1)^{S_C+1}\sqrt{3(2S_A+1)(2S_B+1)(2S_C+1)} \nonumber\\
  &&\times
  \left\{
           \begin{array}{ccc}
                    s_1    & s_2    & S_A \\
                    s_4    & s_3    & 1 \\
                    S_B    & S_C    & S \\
                    \end{array}
     \right\}.
\end{eqnarray}
The $s_i$ denote the spin of $i$-th quark. For an initial $n^1S_0$ ($n\geq2$) strange meson, $S_A=0$.  In contrast, $S_A=1$ for an initial $0^-$ hybrid meson. When the final mesons $B$ and $C$ are two $S$-wave vector mesons ($S_B=S_C=1$, $L_B=L_C=0$), the decay proceeds through $p$-wave ($L_{BC}=1$). Since $\boldsymbol{J}_A=\boldsymbol{S}\oplus\boldsymbol{L}_{BC}$, the allowed value of $S$ is fixed to be 1. Then the $9-j$ Wigner coefficient in Eq. (\ref{eq10}) vanish for the decay of a pseudoscalar hybrid meson into two vector mesons. In contrast, this $9-j$ Wigner coefficient is nonzero for an excited pseudoscalar meson decaying into two vector mesons. Thus, the selection rule for the decays of a pseudoscalar hybrid and conventional mesons is demonstrated.

%


\end{document}